\documentclass[aps,pra,twocolumn,groupedaddress]{revtex4-1}
\usepackage{amsmath}
\usepackage{amssymb}
\usepackage{mathrsfs}
\usepackage[pdftex,colorlinks=true,urlcolor=blue,linkcolor=blue,citecolor=blue]{hyperref}
\usepackage{graphicx}
\usepackage{float}
\usepackage[abs]{overpic}
\usepackage[authormarkup=none,final]{changes}
\definechangesauthor[name={ShovanA},color=blue]{SA}
\definechangesauthor[name={ShovanD},color=red]{SD}

\DeclareMathOperator{\sech}{sech}

\begin{document}

\title{Protocol to engineer Fulde-Ferrell-Larkin-Ovchinnikov states in a cold Fermi gas}

\author{Shovan Dutta}
\author{Erich J. Mueller}
\affiliation{Laboratory of Atomic and Solid State Physics, Cornell University, Ithaca, New York 14853, USA}

\date{\today}

\begin{abstract}
We propose a two-step experimental protocol to directly engineer Fulde-Ferrell-Larkin-Ovchinnikov (FFLO) states in a cold two-component Fermi gas loaded into a quasi-one-dimensional trap. First, one uses phase imprinting to create a train of domain walls in a superfluid with equal number of $\uparrow$- and $\downarrow$-spins. Second, one applies a radio-frequency sweep to selectively break Cooper pairs near the domain walls and transfer the $\uparrow$-spins to a third spin state which does not interact with the $\uparrow$- and $\downarrow$-spins. The resulting FFLO state has exactly one unpaired $\downarrow$-spin in each domain wall and is stable for all values of domain-wall separation and interaction strength. We show that the protocol can be implemented with high fidelity at sufficiently strong interactions for a wide range of parameters available in present-day experimental conditions.
\end{abstract}


\maketitle

\section{\label{intro}Introduction}
Ever since Fulde and Ferrell \cite{fulde1964superconductivity} and Larkin and Ovchinnikov \cite{[{}] [{ [Sov. Phys. JETP {\bf 20}, 762 (1965)]. }]larkin1965inhomogeneous} (FFLO) predicted translational symmetry breaking in superconductors with magnetic impurities, there has been an intense search for physical examples of the phenomenon \cite{casalbuoni2004inhomogeneous}. Although thermodynamic evidence has been found in certain heavy-fermion superconductors \cite{koutroulakis2010field, cho2009upper, kenzelmann2008coupled, miclea2006pressure, kumagai2006fulde, kakuyanagi2005texture, watanabe2004high, capan2004anisotropy, radovan2003magnetic, bianchi2003possible}, layered organic superconductors \cite{koutroulakis2016microscopic, mayaffre2014evidence, agosta2012experimental, uji2012magnetic, bergk2011magnetic, wright2011zeeman, yonezawa2008anomalous, lortz2007calorimetric, uji2006vortex, symington2001evidence, singleton2000observation}, and cold Fermi gases in elongated traps \cite{liao2010spin, revelle20161d}, the phase space for the FFLO state is generically small. As we suggested in a recent Letter \cite{[{}] [{ ({\it accepted in Phys. Rev. Lett.}). }]dutta2016collective}, one can enlarge this parameter space by circumventing thermodynamics, and directly engineering the FFLO state. There we argued that such an engineered FFLO superfluid would be long-lived. Here we give a detailed protocol for this engineering, thereby greatly extending the ability to produce and study the FFLO phase.

\begin{figure}
\includegraphics[width=\columnwidth]{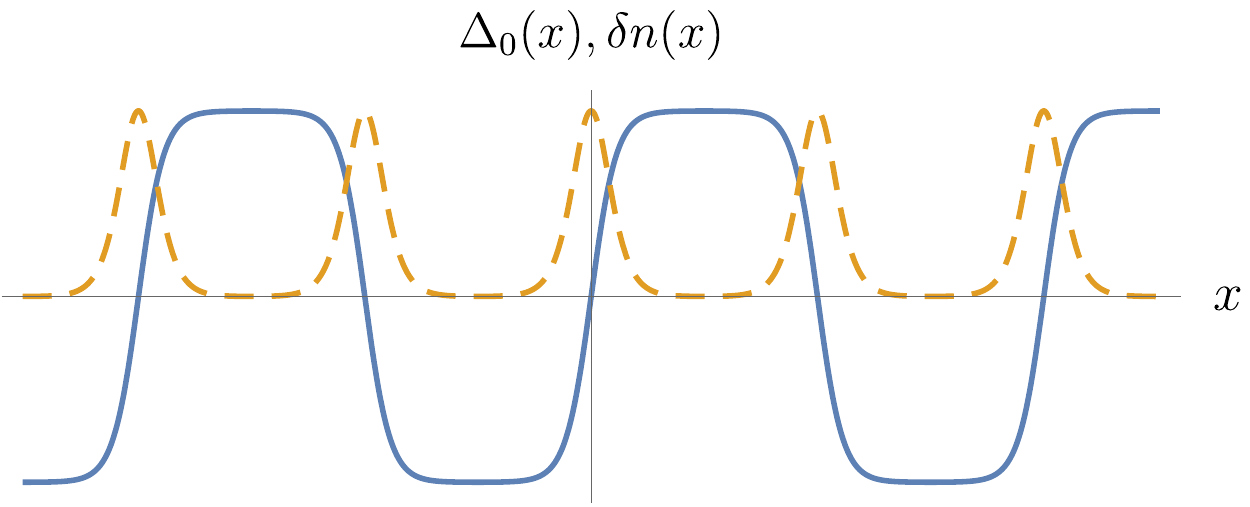}
\caption{\label{fflostate}Spatial variation in the FFLO state. Solid blue curve shows the order parameter or pair wavefunction $\Delta_0(x)$. Dashed red curve shows the density of unpaired fermions, $\delta n(x) \equiv |n_{\uparrow} (x) - n_{\downarrow}(x)|$. The unpaired fermions are localized near the domain walls.}
\end{figure}

In a two-component system of fermions, superconductivity typically occurs when spin-$\uparrow$ particles form Cooper pairs with spin-$\downarrow$ particles. Magnetic impurities can change the relative chemical potentials of the $\uparrow$- and $\downarrow$-spins, breaking pairs and frustrating superconductivity. In cold Fermi gases, where the spin-relaxation time exceeds the timescale of the experiment, similar physics occurs when more $\downarrow$-spins than $\uparrow$-spins are placed in a trap, making an imbalanced (or spin-imbalanced) gas. In 1964, Fulde and Ferrell \cite{fulde1964superconductivity} argued that one could find exotic pairing in such systems, where the Cooper pairs condense into a state with finite momentum, $\Delta_0 (x) \sim e^{i k_0 x}$. At the same time, Larkin and Ovchinnikov \cite{[{}] [{ [Sov. Phys. JETP {\bf 20}, 762 (1965)]. }]larkin1965inhomogeneous} proposed that such systems will have an oscillatory order parameter, $\Delta_0 (x) \sim \cos k_0 x$, an ansatz which is energetically more favorable. Subsequent work found that one generally expects a train of domain walls (solitons), where the order parameter periodically changes sign \cite{[{}] [{ [\href{http://www.jetpletters.ac.ru/ps/1354/article_20458.shtml}{JETP Lett. {\bf 31}, 456 (1980)}].}]brazovskii1980exact, horovitz1981soliton, [{}] [{ [\href{http://www.jetp.ac.ru/cgi-bin/e/index/e/59/2/p434?a=list}{Sov. Phys. JETP {\bf 59}, 434 (1984)}].}]brazovskii1984peierls, mertsching1981incommensurate, machida1984superconductivity, [{}] [{ [\href{http://www.jetp.ac.ru/cgi-bin/e/index/e/58/2/p428?a=list}{Sov. Phys. JETP {\bf 58}, 428 (1983)}]; }]buzdin1983phase, [{}] [{ [\href{http://www.jetp.ac.ru/cgi-bin/e/index/e/66/2/p422?a=list}{Sov. Phys. JETP {\bf 66}, 422 (1987)}]. }]buzdin1987nonuniform}. Larkin and Ovchinnikov's wavefunction is viewed as a special case, where the width of the domain walls is comparable to their separation. In all cases the spin imbalance is concentrated near the order-parameter nodes, where the density of pairs vanishes (Fig.~\ref{fflostate}). These FFLO states have been predicted to occur in a wide range of physical systems, including heavy-fermion superconductors \cite{matsuda2007fulde}, organic supercoductors \cite{beyer2013emerging, shimahara2002fulde, houzet2002new, burkhardt1994fulde}, ultracold Fermi superfluids \cite{radzihovsky2010imbalanced, guan2013fermi, dutta2015dimensional, parish2007quasi, casula2008quantum, baksmaty2011concomitant, zhao2008theory, yang2001inhomogeneous, loh2010detecting, rizzi2008fulde}, and high-density quark matter \cite{sedrakian2009phase, alford2008color, he2006pion, giannakis2005neutral, casalbuoni2004effective, bowers2002crystallography, alford2001crystalline}.

In this paper, we present a simple and robust approach to generating an FFLO state in a superfluid of cold fermionic atoms. We build upon the fact that experimentalists routinely produce superfluids of fermionic lithium or potassium atoms \cite{zwierlein2014superfluidity}, control their environment through optical traps \cite{revelle20161d}, control their spin states with radio waves and microwaves \cite{partridge2006pairing}, and tune their interactions through Feshbach resonance \cite{chin2010feshbach}. After engineering these exotic superfluids, they can probe the order parameter using both in-situ techniques \cite{liao2010spin, revelle20161d} and time-of-flight imaging \cite{yefsah2013heavy, ku2014motion, ku2016cascade}.

Our approach differs from the conventional method of simply cooling an imbalanced gas into the FFLO phase. By coherently driving the system into this state, we overcome the hysteresis and metastability issues which can thwart the traditional prescriptions \cite{baksmaty2011concomitant}.

\section{\label{overview}Overview}
\begin{figure}[h]
\includegraphics[width=0.43\textwidth]{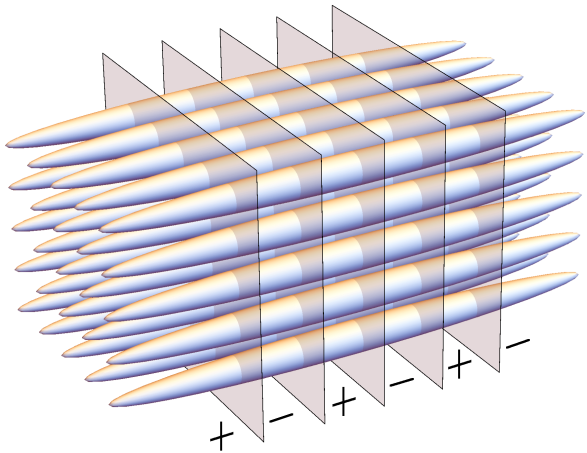}
\caption{\label{balancedprotocolfig}Schematic experimental set-up for producing balanced soliton trains in an array of weakly-coupled tubes. First, uniform superfluids are prepared in each tube by cooling an equal number of attractively interacting $\uparrow$- and $\downarrow$-fermions near a Feshbach resonance. Then solitons (domain walls) are imprinted by shining off-resonant lasers in alternate regions labeled `$-$' to reverse the sign of the local order parameter.}
\end{figure}

We envision a two-component gas of fermionic atoms (two hyperfine states of $^{6}$Li or $^{40}$K) with attractive interactions, loaded into a quasi-one-dimensional (quasi-1D) optical trap consisting of an array of weakly-coupled 1D tubes (Fig.~\ref{balancedprotocolfig}). The 1D nature of each tube leads to Fermi-surface nesting, stabilizing the FFLO states \cite{dutta2015dimensional, parish2007quasi, casula2008quantum, baksmaty2011concomitant}. The small intertube tunneling helps establish long-range superfluid order \cite{zhao2008theory, yang2001inhomogeneous}. To produce FFLO states in each tube, we propose a two-step protocol. In the first step, one creates an array of domain walls (solitons) in a balanced superfluid. To this end, one loads an equal mixture of $\uparrow$- and $\downarrow$-fermions into the trap and cools the system near a Feshbach resonance to form a superfluid, as demonstrated experimentally in \cite{liao2010spin, revelle20161d}. One can create solitons in these superfluids by phase imprinting \cite{ku2016cascade, ku2014motion, yefsah2013heavy, sacha2014proper, law2003dynamic, karpiuk2002solitons, burger2002generation, wu2002controlled, gajda1999optical}, whereby one shines an off-resonant laser pulse on selected portions of the superfluid, which rotates the local phase of the order parameter by $\pi$. Working in a 3D geometry, past experiments \cite{ku2016cascade, ku2014motion, yefsah2013heavy} have demonstrated that one can create solitons in Fermi superfluids by phase imprinting. The same technique has been used extensively in Bose gases \cite{stellmer2008collisions, becker2008oscillations, anderson2001watching, denschlag2000generating, burger1999dark}. A train of solitons can be formed in each tube by imprinting a $\pi$ phase in alternate regions of the trap, as illustrated in Fig.~\ref{balancedprotocolfig}. The tight radial confinement in each tube will prevent the solitons from decaying into vortices and sound waves via the snake instability \cite{ku2016cascade, reichl2017core, bulgac2014quantized, cetoli2013snake, wen2013dark}. This first step is straightforward and we do not model it in detail.

The subject of this paper is analyzing the second step. Once the domain walls (solitons) are formed, we propose using radio waves to selectively break up Cooper pairs in the soliton cores, transferring spin-$\uparrow$ atoms at these locations to a third spin state $|\phi\rangle$ which does not interact with the $|\hspace{-0.1cm}\uparrow\rangle$ and $|\hspace{-0.1cm}\downarrow\rangle$ spin states, thus leaving behind an FFLO state with unpaired $\downarrow$-spins at the nodes. For example, in $^{40}$K one could use \smash{$|\hspace{-0.12cm}\uparrow\rangle = |\frac{9}{2},-\frac{7}{2}\rangle$, $|\hspace{-0.12cm}\downarrow\rangle = |\frac{9}{2},-\frac{9}{2}\rangle$}, and \smash{$|\phi\rangle = |\frac{9}{2},-\frac{5}{2}\rangle$}, where the two numbers denote the total atomic spin $F$ and its projection $m_{\text{F}}$ \cite{sagi2015breakdown}. The frequencies for the atomic transitions are sensitive to the local environment, and, as we will show, one can select frequencies such that the transitions only occur near the cores of the domain walls.

In a recent paper \cite{[{}] [{ ({\it accepted in Phys. Rev. Lett.}). }]dutta2016collective}, we showed that when each soliton in a given tube is filled with exactly one unpaired fermion, the resulting commensurate FFLO (C-FFLO) phase is stable for all values of soliton spacing and interaction strength. In this paper, we will demonstrate that one can produce such long-lived C-FFLO states in a controlled manner by shining radio waves on a balanced soliton train and performing a frequency sweep.

As we describe in Sec.~\ref{spectrum}, a soliton train has gapped bulk modes that are delocalized, and gapless bound states that are localized in the soliton cores. Our protocol utilizes the separation of energy scales between these localized and bulk excitations. The C-FFLO state differs from a balanced soliton train only in the occupation of the bound states. In our protocol we change these occupations by sweeping the energy of radio waves which couple the $|\hspace{-0.1cm}\uparrow\rangle$ and $|\phi\rangle$ states. As in other applications of Rapid Adiabatic Passage ideas \cite{vitanov2001laser, moller2008quantum, rangelov2005stark, malinovsky2001general, vitanov1996landau, rubbmark1981dynamical}, the sweep rate must be slow enough to satisfy adiabaticity. However, the sweep duration is limited by the finite lifetime of the balanced soliton train \cite{[{}] [{ ({\it accepted in Phys. Rev. Lett.}). }]dutta2016collective}. This lifetime increases sharply with interactions. Therefore, one can achieve higher fidelities when the interactions are stronger. Unwanted bulk excitations caused by the sweep can be eliminated by Pauli blocking if one starts with an appropriate density of $|\phi\rangle$-atoms. Even without Pauli blocking, our approach gives relatively few bulk excitations when the bulk gap is large. A larger bulk gap also yields a higher critical temperature \cite{mertsching1981incommensurate, machida1984superconductivity, [{}] [{ [\href{http://www.jetp.ac.ru/cgi-bin/e/index/e/66/2/p422?a=list}{Sov. Phys. JETP {\bf 66}, 422 (1987)}]. }]buzdin1987nonuniform, [{}] [{ [\href{http://www.jetp.ac.ru/cgi-bin/e/index/e/58/2/p428?a=list}{Sov. Phys. JETP {\bf 58}, 428 (1983)}]; }]buzdin1983phase}, thus reducing thermal fluctuations. These arguments further suggest that it is beneficial to work in the strongly interacting regime. We analyze this protocol in detail in Sec.~\ref{CFFLO}, showing that current experiments are in a parameter range where one can generate long-lived C-FFLO states with high fidelity.

Our results are based on a mean-field self-consistent Bogoliubov de-Gennes (BdG) formalism which gives an accurate description of quasi-1D Fermi gases for moderate to weak interactions, and is semiquantitative for stronger interactions \cite{liu2007fulde, liu2008finite, parish2007quasi, mizushima2005direct, baksmaty2011bogoliubov, sun2012oscillatory, radzihovsky2010imbalanced, lutchyn2011spectroscopy, edge2009signature, edge2010collective, xu2014dark, efimkin2015moving}. In addition, past theoretical studies have shown that 1D BdG equations correctly models the equilibrium properties of an array of tubes \cite{[{}] [{ [\href{http://www.jetp.ac.ru/cgi-bin/e/index/e/66/2/p422?a=list}{Sov. Phys. JETP {\bf 66}, 422 (1987)}]. }]buzdin1987nonuniform, [{}] [{ [\href{http://www.jetp.ac.ru/cgi-bin/e/index/e/58/2/p428?a=list}{Sov. Phys. JETP {\bf 58}, 428 (1983)}]; }]buzdin1983phase, lutchyn2011spectroscopy}. As we will show, our protocol depends primarily on a separation of energy scales between the localized and bulk excitations of a soliton train. It is not contingent on the quantitative details.

The rest of the paper is organized as follows. In Sec.~\ref{quasiparticles} we describe the Bogoliubov modes of a train of solitons (or domain walls) and show how the generation of a C-FFLO state from a balanced soliton train is equivalent to changing the mode occupations. In Sec.~\ref{CFFLO}, we model the radio-frequency sweep which implements this change. We carefully analyze different processes that could affect the generation of the C-FFLO state, finding parameter regimes where the protocol has high fidelity. We conclude with a summary and outlook in Sec.~\ref{discussion}.

\section{\label{quasiparticles} Quasiparticle modes}
In this section, we cast the problem of generating the C-FFLO state from a balanced soliton train in terms of the occupation of the Bogoliubov modes. This formalism is convenient for modeling the population transfer by the radio-frequency sweep.

There are at least two competing conventions in the literature for defining the Bogoliubov operators: in the most common one, the spectrum has only positive energies, but there are two types of Bogoliubov modes,  $\hat{\gamma}_j$ and \smash{$\hat{\zeta}_j$}. We use a somewhat less common convention in which the quasiparticle spectrum is symmetric for positive and negative energies, and there is only one type of Bogoliubov mode $\hat{\gamma}_j$. This latter convention is particularly convenient for polarized gases. To avoid any confusion later on, we first summarize both conventions in the next subsection, discussing how they relate to one another. We provided a similar discussion in the Supplemental Material for \cite{[{}] [{ ({\it accepted in Phys. Rev. Lett.}). }]dutta2016collective}.
\subsection{\label{convention}Convention for Bogoliubov operators}
A system of spin-1/2 fermions with short-ranged attractive interactions is described by the Hamiltonian
\begin{align}
\nonumber \hat{H} = \int\hspace{-0.05cm} dx \Big[\hspace{0.1cm}&\sum\nolimits_{\sigma=\uparrow,\downarrow}\hspace{-0.1cm} \hat{\Psi}_{\sigma}^{\dagger}(x) (\hat{H}_0 - \mu_{\sigma}) \hat{\Psi}_{\sigma}(x)\\
& + g_{\mbox{\tiny{1D}}} \hat{\Psi}_{\uparrow}^{\dagger}(x)\hat{\Psi}_{\downarrow}^{\dagger}(x) \hat{\Psi}_{\downarrow}(x) \hat{\Psi}_{\uparrow} (x)\Big],
\label{Hspinhalf}
\end{align}
where $\hat{\Psi}_{\sigma} (x)$ denote the fermion field operators, $\hat{H}_0$ is the single-particle Hamiltonian, $\mu_{\uparrow,\downarrow} \equiv \epsilon_{\mbox{\tiny{F}}} \mp h$ are the chemical potentials of the two spins, $\epsilon_{\mbox{\tiny{F}}}$ being the Fermi energy, and $g_{\mbox{\tiny{1D}}}$ denotes the 1D coupling constant whose relationship with the 3D scattering length is well studied \cite{olshanii1998atomic, bergeman2003atom, haller2010confinement, dutta2015dimensional}. Attractive interactions ($g_{\mbox{\tiny{1D}}} < 0$) lead to Cooper pairing, which gives rise to the superfluid order parameter \smash{$\Delta_0(x) \equiv g_{\mbox{\tiny{1D}}}\langle \hat{\Psi}_{\downarrow} (x) \hat{\Psi}_{\uparrow} (x)\rangle$}. Ignoring quadratic fluctuations about $\Delta_0 (x)$ yields the mean-field Hamiltonian
\begin{align}
\nonumber \hat{H}_{\mbox{\tiny{MF}}} \hspace{-0.03cm}=\hspace{-0.1cm} \int \hspace{-0.1cm} d x\hspace{0.05cm} \bigg[&
\big(\hat{\Psi}^{\dagger}_{\uparrow} \;\;\; \hat{\Psi}_{\downarrow}\big)
\begin{pmatrix} \hat{H}_0 - \mu_{\uparrow} & \Delta_0(x) \\ \Delta_0^* (x) & -\hat{H}_0 + \mu_{\downarrow} \end{pmatrix} 
\bigg(\hspace{-0.1cm}\begin{array}{c} \hat{\Psi}_{\uparrow} \\ \hat{\Psi}^{\dagger}_{\downarrow} \end{array}\hspace{-0.1cm}\bigg)\\
 &- g_{\mbox{\tiny{1D}}}^{-1} \big|\Delta_0(x)\big|^2\bigg] + \text{Tr}\hspace{0.03cm}\big(\hat{H}_0 - \epsilon_{\mbox{\tiny{F}}} - h\big) \hspace{0.05cm},
\label{HMFspinhalf}
\end{align}
where $\hat{\Psi}_{\sigma} \equiv \hat{\Psi}_{\sigma} (x)$. The mean-field Hamiltonian can be diagonalized by solving the BdG equations
\begin{equation}
\begin{pmatrix} \hat{H}_0 - \epsilon_{\mbox{\tiny{F}}} & \Delta_0(x) \\ \Delta_0^* (x) & -\hat{H}_0 + \epsilon_{\mbox{\tiny{F}}} \end{pmatrix} 
\bigg(\hspace{-0.1cm}\begin{array}{c} u_j (x) \\ v_j (x) \end{array}\hspace{-0.1cm}\bigg) = \epsilon_j \bigg(\hspace{-0.1cm}\begin{array}{c} u_j (x) \\ v_j (x) \end{array}\hspace{-0.1cm}\bigg) \hspace{0.02cm},
\label{BdGspinhalf}
\end{equation}
which has a symmetric spectrum: if $(u_j(x)\;\;v_j(x))^{T}$ is an eigenvector with eigenvalue $\epsilon_j$, then $(-v^*_j(x)\;\;u^*_j(x))^T$ is an eigenvector with eigenvalue $-\epsilon_j$. The eigenvectors form an orthonormal set, i.e., \smash{$\int dx \hspace{0.03cm}(u_j^* (x) u_{j^{\prime}}(x) + v_j^* (x) v_{j^{\prime}}(x)) = \delta_{j j^{\prime}}$}.

We define the Bogoliubov operators $\hat{\gamma}_j$ as
\begin{equation}
\bigg(\hspace{-0.1cm}\begin{array}{c} \hat{\Psi}_{\uparrow}(x) \\ \hat{\Psi}^{\dagger}_{\downarrow}(x) \end{array}\hspace{-0.1cm}\bigg) = \sum_j \bigg(\hspace{-0.1cm}\begin{array}{c} u_j (x) \\ v_j (x) \end{array}\hspace{-0.1cm}\bigg) \hat{\gamma}_j \hspace{0.05cm},
\label{ourdefinition}
\end{equation}
where the sum is over both positive and negative energies. The orthonormality of the eigenvectors ensures that the modes $\hat{\gamma}_j$ are fermionic, i.e., \smash{$\{\hat{\gamma}_j,\hat{\gamma}^{\dagger}_{j^{\prime}}\} = \delta_{j j^{\prime}}$}. Substituting Eq.~\eqref{ourdefinition} into Eq.~\eqref{HMFspinhalf} we find
\begin{equation}
\hat{H}_{\mbox{\tiny{MF}}} \hspace{-0.03cm}= \sum_j (\epsilon_j + h) \hat{\gamma}^{\dagger}_j \hat{\gamma}_j + \text{Tr}\hspace{0.03cm}\big(\hat{H}_0 - \epsilon_{\mbox{\tiny{F}}} - h\big) - g_{\mbox{\tiny{1D}}}^{-1} \hspace{-0.1cm}\int \hspace{-0.1cm} dx \big|\Delta_0(x)\big|^2 .
\label{HMFourconvention}
\end{equation}
The occupation of the modes is given by $\smash{\langle \hat{\gamma}^{\dagger}_j \hat{\gamma}_j \rangle} = n_{\mbox{\tiny{F}}}(\epsilon_j + h)$ where $n_{\mbox{\tiny{F}}}$ denotes the Fermi function. Thus at zero temperature, all quasiparticle modes with energy $\epsilon_j < -h$ are occupied, and all other modes are empty. In particular, for $h\equiv \mu_{\downarrow} - \mu_{\uparrow} =0$ (no spin imbalance), all negative energy modes are occupied and positive energy modes are empty. When $\mu_{\downarrow} > \mu_{\uparrow}$ ($h>0$), one has to remove quasiparticles from the modes with energy between 0 and $-h$, resulting in a net excess of $\downarrow$-spins. Similarly, if $\mu_{\uparrow} > \mu_{\downarrow}$, one populates the modes between 0 and $|h|$, resulting in a net excess of $\uparrow$-spins.

One arrives at the other convention by noting that Eq.~\eqref{ourdefinition} can be written as
\begin{align}
\nonumber \bigg(\hspace{-0.1cm}\begin{array}{c} \hat{\Psi}_{\uparrow}(x) \\ \hat{\Psi}^{\dagger}_{\downarrow}(x)  \end{array}\hspace{-0.1cm}\bigg) &= \sum_{\epsilon_j>0} \bigg(\hspace{-0.1cm}\begin{array}{c} u_j(x)   \\ v_j(x)   \end{array}\hspace{-0.1cm}\bigg) \hat{\gamma}_j + \sum_{\epsilon_j<0} \bigg(\hspace{-0.1cm}\begin{array}{c} u_j(x)   \\ v_j(x)   \end{array}\hspace{-0.1cm}\bigg) \hat{\gamma}_j \\
\nonumber & = \sum_{\epsilon_j>0} \bigg(\hspace{-0.1cm}\begin{array}{c} u_j (x)  \\ v_j (x)  \end{array}\hspace{-0.1cm}\bigg) \hat{\gamma}_j + \sum_{\epsilon_j>0} \bigg(\hspace{-0.1cm}\begin{array}{c} -v^*_j(x)   \\ u^*_j(x)  \end{array}\hspace{-0.1cm}\bigg) \hat{\zeta}_j^{\dagger} \\
& = \sum_{\epsilon_j>0} \begin{pmatrix} u_j(x)   & -v^*_j(x)   \\ v_j(x)  & u^*_j(x)  \end{pmatrix}
 \bigg(\hspace{-0.1cm}\begin{array}{c} \hat{\gamma}_j \\ \hat{\zeta}_j^{\dagger} \end{array}\hspace{-0.1cm}\bigg) \hspace{0.05cm},
\label{otherdefinition}
\end{align}
where we have used the fact that for each state $(u_j\;\;v_j)^T$ with energy $\epsilon_j$, there is a state \smash{$(-v_j^*\;\;u_j^*)^T$} with energy $-\epsilon_j$, and defined new fermionic operators \smash{$\hat{\zeta}_j \equiv \hat{\gamma}_j^{\dagger}$} for $\epsilon_j <0$. The operators $\hat{\gamma}_j$ and \smash{$\hat{\zeta}_j$} in Eq.~\eqref{otherdefinition} represent the Bogoliubov modes in this other convention. Substituting Eq.~\eqref{otherdefinition} into Eq.~\eqref{HMFspinhalf}, we obtain
\begin{align}
\nonumber \hat{H}_{\mbox{\tiny{MF}}} \hspace{-0.03cm}= &\sum_{\epsilon_j>0} \big[(\epsilon_j + h) \hat{\gamma}^{\dagger}_j \hat{\gamma}_j + (\epsilon_j - h)\hspace{0.03cm} \hat{\zeta}^{\dagger}_j\hat{\zeta}_j - (\epsilon_j - h) \big]  \\
& + \text{Tr}\hspace{0.03cm}\big(\hat{H}_0 - \epsilon_{\mbox{\tiny{F}}} - h\big) - g_{\mbox{\tiny{1D}}}^{-1} \int\hspace{-0.05cm} dx \hspace{0.05cm}\big|\Delta_0(x)\big|^2 \hspace{0.05cm}.
\label{HMFotherconvention}
\end{align}
The occupations of the modes are given by $\langle \hat{\gamma}^{\dagger}_j \hat{\gamma}_j \rangle = n_{\mbox{\tiny{F}}}(\epsilon_j + h)$ and \smash{$\langle \hat{\zeta}^{\dagger}_j \hat{\zeta}_j \rangle = n_{\mbox{\tiny{F}}}(\epsilon_j - h)$}. At zero temperature, only the $\hat{\gamma}$ modes with $\epsilon_j < -h$ and the \smash{$\hat{\zeta}$} modes with $\epsilon_j < h$ are occupied. However $\epsilon_j >0$, so there are no negative energy modes, which means in the balanced case ($h=0$), all Bogoliubov modes are empty. Excess $\downarrow$-spins ($h>0$) are incorporated by filling up only the \smash{$\hat{\zeta}$} modes with $0<\epsilon_j < h$, whereas excess $\uparrow$-spins ($h<0$) are incorporated by filling up only the \smash{$\hat{\gamma}$} modes with $0<\epsilon_j < |h|$. Although the two conventions yield different descriptions of a state, they are formally equivalent.

\subsection{\label{spectrum}Quasiparticle spectrum of a soliton train}
Here we summarize a few important features of the fermionic quasiparticle spectrum of a soliton train that are relevant for analyzing the effect of a radio-frequency sweep. We also establish the connection between the occupation of the Bogoliubov modes with the generation of C-FFLO states. Further details on the spectrum of a soliton train can be found in \cite{[{}] [{ ({\it accepted in Phys. Rev. Lett.}). }]dutta2016collective, mertsching1981incommensurate, machida1984superconductivity, [{}] [{ [\href{http://www.jetp.ac.ru/cgi-bin/e/index/e/66/2/p422?a=list}{Sov. Phys. JETP {\bf 66}, 422 (1987)}]. }]buzdin1987nonuniform, [{}] [{ [\href{http://www.jetp.ac.ru/cgi-bin/e/index/e/58/2/p428?a=list}{Sov. Phys. JETP {\bf 58}, 428 (1983)}]; }]buzdin1983phase, [{}] [{ [\href{http://www.jetpletters.ac.ru/ps/1354/article_20458.shtml}{JETP Lett. {\bf 31}, 456 (1980)}].}]brazovskii1980exact, horovitz1981soliton, [{}] [{ [\href{http://www.jetp.ac.ru/cgi-bin/e/index/e/59/2/p434?a=list}{Sov. Phys. JETP {\bf 59}, 434 (1984)}].}]brazovskii1984peierls}.

Following our approach in \cite{[{}] [{ ({\it accepted in Phys. Rev. Lett.}). }]dutta2016collective} we use the Andreev approximation~\cite{[{}] [{ [\href{http://www.jetp.ac.ru/cgi-bin/e/index/e/19/5/p1228?a=list}{Sov. Phys. JETP {\bf 19}, 1228 (1964)}].}]andreev1964thermal}, whereby one linearizes the dispersion about the Fermi points and considers right-moving and left-moving modes separately. With this approximation one can solve the BdG equations analytically, which is particularly useful to obtain a qualitative understanding of the physics and estimating the variation of physical quantites, such as the bulk gap, with the experimental parameters. However, this approximation is strictly valid only for weak interactions where pairing is limited to the vicinity of each Fermi point. As we will see in Sec.~\ref{CFFLO}, the validity of our protocol does not depend on making the Andreev approximation. It only rests on a few generic features, such as a separation of energy scales between localized and bulk excitations, that are also present in the full model. We will only use the Andreev approximation to estimate the range of parameters over which the protocol has high fidelity. We find good numerical agreement of these estimates with the full BdG equations.

The BdG equations [Eq.~\eqref{BdGspinhalf}] for the coherence factors in a 1D tube can be expressed as
\begin{equation}
\begin{pmatrix} -\partial_x^2 /2 - \epsilon_{\mbox{\tiny{F}}} & \Delta_0(x) \\ \Delta_0^* (x) & \partial_x^2 /2 + \epsilon_{\mbox{\tiny{F}}} \end{pmatrix} 
\bigg(\hspace{-0.1cm}\begin{array}{c} u_j (x) \\ v_j (x) \end{array}\hspace{-0.1cm}\bigg) = \epsilon_j \bigg(\hspace{-0.1cm}\begin{array}{c} u_j (x) \\ v_j (x) \end{array}\hspace{-0.1cm}\bigg) \hspace{0.02cm},
\label{BdGsolitontrain}
\end{equation}
where we have set $\hbar = m = 1$, $m$ being the mass of each fermion. For sufficiently weak interactions, only the modes near the Fermi points contribute to pairing. Thus, as already explained, we make the Andreev approximation, where we write the fermion fields as a sum over right-moving and left-moving Bogoliubov modes $\hat{\gamma}_j^{\pm}$ [see Eq.~\eqref{ourdefinition}],
\begin{equation}
\bigg(\hspace{-0.1cm}\begin{array}{c} \hat{\Psi}_{\uparrow}(x) \\ \hat{\Psi}^{\dagger}_{\downarrow}(x) \end{array}\hspace{-0.1cm}\bigg) = \sum_{s=\pm,j} e^{i s k_{\mbox{\tiny{F}}} x} \bigg(\hspace{-0.1cm}\begin{array}{c} u_j^{s} (x) \\ v_j^s (x) \end{array}\hspace{-0.1cm}\bigg) \hat{\gamma}_j^s \hspace{0.05cm},
\label{rightleft}
\end{equation}
where
\begin{equation}
\Big(\hspace{-0.03cm}-\frac{\partial_x^2}{2} - \epsilon_{\text{F}}\hspace{-0.03cm}\Big)\hspace{-0.03cm}\bigg[\bigg(\hspace{-0.1cm}\begin{array}{c} u_j^{\pm}(x) \\ v_j^{\pm}(x) \end{array}\hspace{-0.1cm}\bigg) e^{\pm i k_{\text{F}} x}\bigg] \hspace{-0.07cm}\approx\hspace{-0.07cm} \bigg[\hspace{-0.02cm}\mp i k_{\text{F}} \partial_x \bigg(\hspace{-0.1cm}\begin{array}{c} u_j^{\pm}(x) \\ v_j^{\pm}(x) \end{array}\hspace{-0.1cm}\bigg)\hspace{-0.05cm}\bigg] e^{\pm i k_{\text{F}} x}
\label{andreev}
\end{equation}
and $\{\hat{\gamma}_j^s,\hat{\gamma}^{s^{\prime}\dagger}_{j^{\prime}}\} = \delta_{s s^{\prime}}\delta_{j j^{\prime}}$ where $k_{\mbox{\tiny{F}}}$ is the Fermi momentum. The BdG equations for the right-moving and left-moving Bogoliubov modes can be obtained be substituting $(u_j(x),v_j(x)) = (u_j^{\pm}(x),v_j^{\pm}(x)) \hspace{0.05cm} e^{\pm i k_{\mbox{\tiny{F}}} x}$ in Eq.~\eqref{BdGsolitontrain} and using Eq.~\eqref{andreev}, which yield
\begin{align}
\begin{pmatrix} \mp i k_{\text{F}} \partial_x & \Delta_0(x) \\ \Delta_0^*(x) & \pm i k_{\text{F}}\partial_x \end{pmatrix} 
\left(\hspace{-0.1cm}\begin{array}{c} u^{\pm}_j(x) \\ v^{\pm}_j(x) \end{array}\hspace{-0.1cm}\right) = \epsilon^{\pm}_j \left(\hspace{-0.1cm}\begin{array}{c} u^{\pm}_j(x) \\ v^{\pm}_j(x) \end{array}\hspace{-0.1cm}\right) ,
\label{andreevBdG}\\
\text{where} \;\;\Delta_0(x) = g_{\mbox{\tiny{1D}}} \sum_{s=\pm,j} n_{\mbox{\tiny{F}}} (\epsilon^s_j + h)\hspace{0.05cm} u^{s}_j(x) v^{s *}_j(x) \hspace{0.05cm}.
\label{andreevselfconsistency1}
\end{align}
For real $\Delta_0 (x)$, the right- and left-moving branches are related by a complex conjugation: $(u^-,v^-) = (u^+,v^+)^*$ and $\epsilon^- = \epsilon^+=\epsilon$. Thus we can rewrite Eq.~\eqref{andreevselfconsistency1} as
\begin{equation}
\Delta_0(x) = 2g_{\mbox{\tiny{1D}}} \sum\nolimits_j n_{\mbox{\tiny{F}}} (\epsilon_j + h)\hspace{0.05cm} \text{Re}\hspace{-0.05cm}\left[ u_j^+(x) v^{+*}_j(x)\right] .
\label{andreevselfconsistency2}
\end{equation}

A periodic solution to Eqs.~\eqref{andreevBdG} and \eqref{andreevselfconsistency2} has the soliton train profile $\Delta_0 (x) = \Delta_1 k_1 \text{sn}(\Delta_1 x/k_{\mbox{\tiny{F}}}, k_1)$, where $\Delta_1 = 2 k_{\mbox{\tiny{F}}} k_0 K(k_1)/\pi$ \cite{[{}] [{ [\href{http://www.jetpletters.ac.ru/ps/1354/article_20458.shtml}{JETP Lett. {\bf 31}, 456 (1980)}].}]brazovskii1980exact, horovitz1981soliton, [{}] [{ [\href{http://www.jetp.ac.ru/cgi-bin/e/index/e/59/2/p434?a=list}{Sov. Phys. JETP {\bf 59}, 434 (1984)}].}]brazovskii1984peierls, mertsching1981incommensurate, machida1984superconductivity, [{}] [{ [\href{http://www.jetp.ac.ru/cgi-bin/e/index/e/66/2/p422?a=list}{Sov. Phys. JETP {\bf 66}, 422 (1987)}]. }]buzdin1987nonuniform, [{}] [{ [\href{http://www.jetp.ac.ru/cgi-bin/e/index/e/58/2/p428?a=list}{Sov. Phys. JETP {\bf 58}, 428 (1983)}]; }]buzdin1983phase}. Here $2\pi/k_0$ denotes the period, sn is a Jacobi elliptic function \cite{whittaker1996course}, $K$ denotes the complete elliptic integral of the first kind, and $k_1 \in (0,1)$ parametrizes the sharpness of each soliton. The modes are characterized by the parameter $k_1$ which is in turn set by the self-consistency condition in Eq.~\eqref{andreevselfconsistency2}. Many of our results are conveniently expressed in terms of $w \equiv (k_0/k_{\mbox{\tiny{F}}}) \exp(\pi k_{\mbox{\tiny{F}}} a_{\mbox{\tiny{1D}}}/2)$ which corresponds to the width of each soliton in units of the separation between solitons ($k_0^{-1}$). This ratio quantifies the effects of interactions in a soliton train. The width of a soliton is determined by the interaction strength, and for fixed $k_0$, decreasing the interactions increases $w$. If the interactions become too weak, the superfluid becomes too frail to support the soliton train, and the system is driven normal. Thus for balanced soliton trains ($h=0$) one must have $w \lesssim 4$. For $w \lesssim 1$, one enters the strongly interacting regime.

\begin{figure*}
\includegraphics[scale=1]{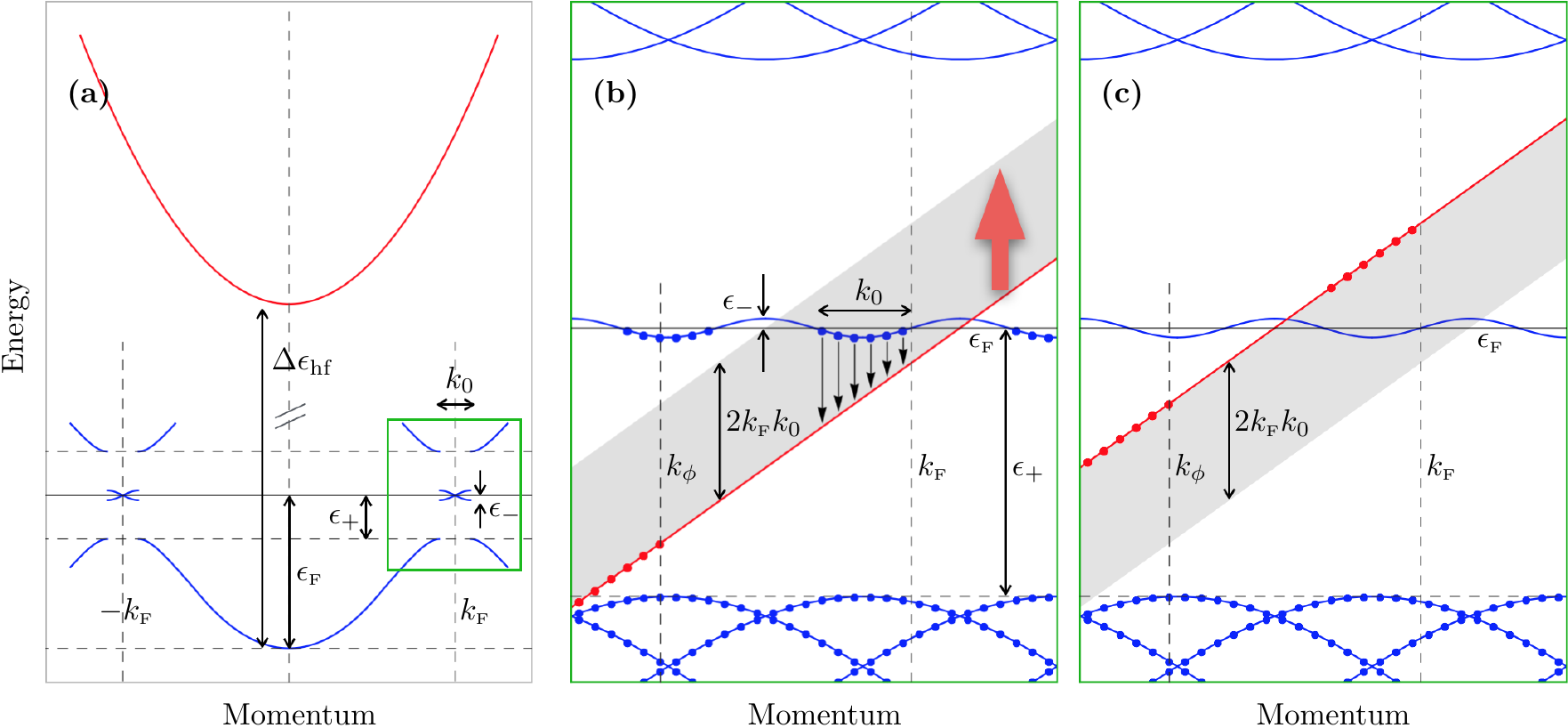}
\caption{\label{cffloprotocolfig}{\bf (a).} Blue curves: Bogoliubov spectrum of a soliton train, containing bulk modes with $|\epsilon|>\epsilon_+$ and midgap modes with $|\epsilon|<\epsilon_-$, where $\epsilon$ is measured from the Fermi level $\epsilon_{\mbox{\tiny{F}}}$. Red curve: quadratic dispersion of the noninteracting spin state $|\phi\rangle$. The internal energy difference $\Delta\epsilon_{\text{hf}} \gg \epsilon_{\mbox{\tiny{F}}}$. {\bf (b)} and {\bf (c).} Blue curves: right-moving Bogoliubov modes [boxed in {\bf (a)}] in the repeated-zone representation. Red line: $|\phi\rangle$ dispersion shifted by the radio frequency $\omega$, which is swept from $\omega_{\text{max}} = \Delta\epsilon_{\text{hf}} + k_{\mbox{\tiny{F}}} k_0/2$ [in {\bf (b)}] to $\omega_{\text{min}} = \Delta\epsilon_{\text{hf}} - 3 k_{\mbox{\tiny{F}}} k_0/2$ [in {\bf (c)}]. Small dots indicate occupied states. The RF sweep couples a filled quasiparticle state with an empty $|\phi\rangle$ state at the same momentum, and {\it vice versa}. If the coupling is sufficiently adiabatic (see Sec.~\ref{adiabaticity}), the sweep will transfer all particles from the midgap modes to the resonantly coupled $|\phi\rangle$ states [as in {\bf (c)}]. If all $|\phi\rangle$ states below a suitable momentum \smash{$k_{\phi} \equiv \sqrt{2 \epsilon_{\phi}}$} are initially occupied, the sweep does not affect the bulk modes or the vacant midgap modes.}
\end{figure*}

Since $\Delta_0 (x)$ is periodic, each Bogoliubov mode can be labeled by a quasimomentum lying in the first Brillouin zone. The energy spectrum $\epsilon(k)$ is most conveniently expressed in the extended zone representation as
\begin{equation}
\frac{k}{k_0} = \pm \frac{1}{\pi} \frac{\epsilon}{\epsilon_+} \text{Re} \hspace{-0.05cm} \left[ \sqrt{\frac{\epsilon_-^2 - \epsilon^2}{\epsilon_+^2 - \epsilon^2}} \hspace{0.1cm}\Pi \hspace{-0.05cm} \left(\frac{\epsilon_+^2 - \epsilon_-^2}{\epsilon_+^2 - \epsilon^2}, \sqrt{1-\frac{\epsilon_-^2}{\epsilon_+^2}}\right) \right],
\label{quasiparticlespectrum}
\end{equation}
where $\epsilon_{\pm} \equiv \frac{1}{2}(1 \pm k_1) \Delta_1$ and $\Pi$ denotes the complete elliptic integral of the third kind. As per our convention, the spectrum is symmetric for positive and negative energies. It has a continuum of bulk modes with $|\epsilon| > \epsilon_+$ and a band of midgap modes for with $|\epsilon| < \epsilon_-$, as seen in the boxed region of Fig.~\ref{cffloprotocolfig}(a). Describing the region outside the box requires going beyond the Andreev approximation. Those modes are not relevant to the processes which we are studying. For sufficiently strong interactions ($w \lesssim 2$), $\epsilon_+ \approx 4 k_{\mbox{\tiny{F}}} k_0 / w$ and $\epsilon_- \approx 16 k_{\mbox{\tiny{F}}} k_0 w^{-1} e^{-4 \pi / w} \ll \epsilon_+$. Hence, the bulk gap increases as $1/w$.

The mode wavefunctions are of the Bloch form, labeled by a quasimomentum $p \in [-k_0/2, k_0/2)$ and an energy $\epsilon$. The positive and negative energy modes are related by a particle-hole transformation: $(u(x),v(x)) \leftrightarrow (-v(x),u(x))$. In addition, one has the symmetry $(u^{\pm}_{-p}(x),v^{\pm}_{-p}(x)) = (v^{\pm *}_{p}(x),u^{\pm *}_{p}(x))$ for modes with the same energy. The midgap modes represent Andreev bound states which are localized in the soliton cores \cite{antezza2007dark, scott2012decay}. For strong enough interactions ($w \lesssim 3$), they are given by (for $\epsilon, p>0$)
\begin{equation}
\hspace{-0.2cm}\left(\hspace{-0.1cm}\begin{array}{c} u^+_p(x) \vspace{0.1cm}\\ v^+_p(x) \end{array}\hspace{-0.1cm}\right) \approx  \sqrt{\frac{\xi}{L}} \left(\hspace{-0.1cm}\begin{array}{c} \sum_{n \text{ even}}\vspace{0.15cm} \\ -i\sum_{n \text{ odd}} \end{array}\hspace{-0.1cm}\right) \frac{e^{i(p+ n k_0) x}}{\cosh \big(2 \xi (n + \frac{p}{k_0})\big)} \hspace{0.05cm},
\label{boundstatesuv}
\end{equation}
where $L$ denotes the length of the system, and $\xi = \pi w/ 16$ represents the width of a bound state around a soliton core: to a good approximation, $|u^+_p(x)|^2,|v^+_p(x)|^2 \propto \smash{\exp(-(k_0 x / \sqrt{2} \xi)^2)}$ for $|k_0 x| < \pi/2$.

The higher-energy bulk modes are relatively unaffected by pairing. Hence, they are well described by plane waves. The lowest-energy bulk mode ($|\epsilon| = \epsilon_+$) is the one most affected. For $w \lesssim 2$, this mode is given by (for $\epsilon,p >0$)
\begin{equation}
\hspace{-0.2cm}\left(\hspace{-0.1cm}\begin{array}{c} u^+(x) \vspace{0.1cm}\\ v^+(x) \end{array}\hspace{-0.1cm}\right) \approx  \frac{w}{8\sqrt{L}} \left(\hspace{-0.1cm}\begin{array}{c} \sum_{n \text{ even}}\vspace{0.15cm} \\ -i\sum_{n \text{ odd}} \end{array}\hspace{-0.1cm}\right) \frac{e^{i(n + 1/2)k_0 x}}{\sinh \big(\frac{\pi w}{8} (n + \frac{1}{2})\big)} \hspace{0.05cm}.
\label{freestatesuv}
\end{equation}

Note that the coherence factors $u_p^+(x)$ [and $v_p^+(x)$] for both midgap modes and bulk modes can be written in the form \smash{$u^+_p(x) = (1/\sqrt{L}) \sum_n \bar{u}^j_{p,n} \hspace{0.02cm} e^{i(p + 2 n k_0) x}$} where $n$ is an integer and $-k_0\leq p <k_0$. This is because the soliton train has an additional symmetry, $\Delta_0 (x + \pi/k_0) = -\Delta_0 (x)$, which decouples the even and odd Fourier modes in the BdG equations, effectively doubling the size of the Brillouin zone \cite{[{}] [{ ({\it accepted in Phys. Rev. Lett.}). }]dutta2016collective, edge2009signature}.

For $h \neq 0$, the number of excess fermions per soliton $n_s$ is simply related to the spectrum $\epsilon(k)$ in Eq.~\eqref{quasiparticlespectrum} as $n_s = 2 |k_h| / k_0$ where $\epsilon(k_h) = h$. Hence, the C-FFLO state, with $n_s = 1$, is formed when $\epsilon_- < |h|< \epsilon_+$ [see Fig.~\ref{cffloprotocolfig}(a)]; i.e., when the chemical potentials lie in the gap between bulk modes and midgap modes. Since $\mu_{\uparrow,\downarrow} \equiv \epsilon_{\mbox{\tiny{F}}} \mp h$, a C-FFLO state with excess $\downarrow$-spins is formed when $\epsilon_- < h< \epsilon_+$, whereas the one with excess $\uparrow$-spins is formed when $-\epsilon_+ < h< -\epsilon_-$.

In our convention, detailed in Sec.~\ref{convention}, the occupation of a Bogoliubov mode $\hat{\gamma}_j$ at zero temperature is given by $\smash{\langle \hat{\gamma}^{\dagger}_j \hat{\gamma}_j \rangle} = \Theta(-\epsilon_j - h)$, where $\Theta$ denotes the unit step function. Therefore, a balanced soliton train ($h=0$) is formed by filling up all the negative energy modes. In a C-FFLO state with excess $\downarrow$-spins, only the negative energy bulk modes, with $\epsilon < -\epsilon_+$, are occupied. Therefore, one can produce such a state by vacating all the occupied midgap modes in a balanced soliton train. Conversely, to produce a C-FFLO state with excess $\uparrow$-spins, one needs to fill all the unoccupied midgap modes. This change of occupation can be achieved by a radio-frequency sweep, which we model in the next section.

\section{\label{CFFLO}Generation of C-FFLO states}
Here we model the process of generating a C-FFLO state from a balanced soliton train by a radio-frequency sweep. We describe in detail the physics behind the protocol in Sec.~\ref{modeling}. In Sec.~\ref{adiabaticity} through \ref{typicalparameters} we explore various processes which could interfere with producing the FFLO state, explaining how to choose parameters. We show that the protocol can be implemented with high fidelity in present-day experimental conditions.

Our strategy is to use radio waves to selectively break up pairs in the soliton cores, and convert the spin-$\uparrow$ atoms to a third noninteracting spin state $|\phi\rangle$. As we described in the last section, a balanced soliton train differs from a C-FFLO state in the occupation of the Bogoliubov modes. In our convention the quasiparticle spectrum of a soliton train is symmetric for positive and negative energies, with delocalized bulk modes for $|\epsilon|>\epsilon_+$ and localized midgap modes for $|\epsilon|<\epsilon_-$ [Fig.~\ref{cffloprotocolfig}(a)]. All negative energy modes are occupied in a balanced soliton train. The C-FFLO state with excess $\downarrow$-spins is formed by removing all quasiparticles from the midgap modes. Our key idea is to use a Rapid Adiabatic Passage protocol which uses a radio-frequency (RF) sweep to vacate these midgap modes by transferring the spin-$\uparrow$ atoms to the $|\phi\rangle$ state. A preformed Fermi sea of $|\phi\rangle$-atoms prevents any bulk excitation, though even without the Fermi sea the number of bulk excitations can be small.

\subsubsection{\label{modeling}Modeling the radio-frequency sweep}
We model the coupling of the atoms to radio waves by
\begin{equation}
\hat{H}_{\text{RF}} = \Omega\hspace{-0.05cm} \int \hspace{-0.1cm} dx \hspace{0.08cm}\hat{\Phi}^{\dagger} (x) \hspace{0.03cm}\hat{\Psi}_{\uparrow} (x) \hspace{0.05cm} e^{-i\hspace{-0.05cm}\int\hspace{-0.08cm} dt \hspace{0.03cm}\omega(t)} + \text{h.c.}\hspace{0.05cm},
\label{HRF}
\end{equation}
where $\hat{\Phi}^{\dagger} (x)$ creates a fermion at position $x$ in the spin state $|\phi\rangle$, $\Omega$ is the coupling strength, and $\omega(t)$ is the frequency of the radio waves. In our protocol one sweeps $\omega$ over a small frequency range (a few kHz) around $\Delta \epsilon_{\text{hf}}$, where $\Delta \epsilon_{\text{hf}}$ is the internal energy difference of the $|\hspace{-0.12cm}\uparrow\rangle$ and $|\phi\rangle$ states (hundreds of MHz). Throughout the sweep the coupling is far-off-resonant for the spin-$\downarrow$ atoms. We can write Eq.~\eqref{HRF} in terms of the Bogoliubov operators $\hat{\gamma}_j$ as [see Eq.~\eqref{ourdefinition}]
\begin{equation}
\hat{H}_{\text{RF}} = \Omega\hspace{0.02cm}\sum_j \int \hspace{-0.1cm} dx \hspace{0.08cm\hat{\Phi}^{\dagger} (x) \hspace{0.03cm}}u_j (x) \hspace{0.03cm}\hat{\gamma}_j\hspace{0.05cm} e^{-i\hspace{-0.05cm}\int\hspace{-0.08cm} dt \hspace{0.03cm}\omega(t)} + \text{h.c.}\hspace{0.05cm}.
\label{HRFgamma}
\end{equation}
As can be seen, within our convention, the RF coupling removes quasiparticles from the superfluid while creating particles in the $|\phi\rangle$ state, and {\it vice-versa}. There are right-moving and left-moving Bogoliubov modes centered at the two Fermi points [Fig.~\ref{cffloprotocolfig}(a)]. They respond equally to the applied field, so we will only consider the right-moving modes. As we showed in Sec.~\ref{spectrum}, each right-moving mode can be labeled by a quasimomentum $p \in [-k_0,k_0)$, and an energy $\epsilon$ indexed by $j$, with wavefunctions of the form \smash{$u^j_p(x) = (1/\sqrt{L}) \sum_n \bar{u}^j_{p,n} \hspace{0.02cm} e^{i(k_{\mbox{\tiny{F}}} + p + 2 n k_0) x}$} where $n$ is an integer. The noninteracting state $|\phi\rangle$ is composed of plane-wave eigenstates, \smash{$\hat{\Phi}^{\dagger}(x) = (1/\sqrt{L}) \sum_k e^{-i k x} \hat{\phi}^{\dagger}_k$}. As a result, radio waves only couple $|\phi\rangle$ states with momentum $k_{\mbox{\tiny{F}}} + k$ to Bogoliubov modes with quasimomentum $p_k$ such that $p_k + 2 n_k k_0 = k$ for some integer $n_k$, or $p_k = k - 2 k_0 \lfloor k/2 k_0 + 1/2\rfloor$. Thus we can rewrite Eq.~\eqref{HRFgamma} as
\begin{equation}
\hat{H}_{\text{RF}}^{+} = \Omega\hspace{0.02cm}\sum_k \sum_j \bar{u}^{j}_{k}\hspace{0.05cm} \hat{\phi}^{\dagger}_{k_{\mbox{\tiny{F}}} + k}\hspace{0.05cm} \hat{\gamma}^{j}_{p_k} \hspace{0.05cm} e^{-i\hspace{-0.05cm}\int\hspace{-0.08cm} dt \hspace{0.03cm}\omega(t)} + \text{h.c.}\hspace{0.05cm},
\label{HRFplus}
\end{equation}
where the superscript `+' indicates that we are working with the right-moving Bogoliubov modes, the $k$-summation is over all momenta, the $j$-summation is over different modes with the same quasimomentum $p_k$, and $\bar{u}^{j}_{k} \equiv \bar{u}^{j}_{p_k,n_k}$.

The effect of the coupling in Eq.~\eqref{HRFplus} is best understood in a repeated-zone representation of the Bogoliubov modes. This is shown by the blue curves in Fig.~\ref{cffloprotocolfig}(b) where we also plot the spectrum of the $|\phi\rangle$ states shifted down by $\omega$ (red curve). Near the Fermi point, the $|\phi\rangle$ spectrum is linear with slope $k_{\mbox{\tiny{F}}}$. In this repeated-zone picture, a $|\phi\rangle$ state is coupled to the quasiparticle states at the same momentum, and the coupling is on resonance where the red curve intersects a blue curve. In the experiment, one sweeps $\omega$ over a small range from $\omega_{\text{max}}$ to $\omega_{\text{min}}$ such that all the occupied midgap modes come on resonance at least once, as in Fig.~\ref{cffloprotocolfig}(b). If the sweep is sufficiently adiabatic, the RF coupling will vacate these modes, while populating the resonantly coupled $|\phi\rangle$ states [Fig.~\ref{cffloprotocolfig}(c)]. Since the $|\phi\rangle$-atoms are noninteracting, or very weakly interacting, their momentum distribution cannot change appreciably over the sweep duration, so there is no possibility of refilling any of the unoccupied midgap modes.

There may also be transitions from the bulk modes. These unwanted transitions can be entirely eliminated if all of the $|\phi\rangle$ states below an energy threshold $\epsilon_{\phi}$ are initially occupied, so that the available $|\phi\rangle$ states are far-off-resonant with the bulk modes [Fig.~\ref{cffloprotocolfig}(b)]. As we show below, this can be achieved for a wide range of parameters. Alternatively, if it is inconvenient to pre-fill the trap with $|\phi\rangle$-atoms, one can tune the sweep rate so that it is adiabatic for the midgap modes, but diabatic for the bulk modes, thus only causing transition from the midgap modes. This latter approach requires a separation of scales in the coherence factors $|\bar{u}^{j}_{k}|$ in Eq.~\eqref{HRFplus} between the bulk and the midgap modes. As we will show, this separation of scales exists and becomes larger at stronger interactions.

\subsubsection{\label{adiabaticity}Adiabaticity requirements for midgap modes}
We model a linear frequency sweep with $\omega$ decreasing from $\omega_{\text{max}} = \Delta \epsilon_{\text{hf}} + k_{\mbox{\tiny{F}}} k_0/2$ to $\omega_{\text{min}} = \Delta \epsilon_{\text{hf}} - 3 k_{\mbox{\tiny{F}}} k_0/2$ at a rate $\nu$, as depicted in Figs.~\ref{cffloprotocolfig}(b) and \ref{cffloprotocolfig}(c). During this evolution, each midgap mode is swept through resonance with a $|\phi\rangle$ state with $k \in [-k_0,0]$. The coherence factors $|\bar{u}^{g}_{k}|$ (`g' refers to midgap modes) in this interval are larger than those in any other interval. For each value of $k \in [-k_0,0]$, the RF coupling in Eq.~\eqref{HRFplus} is well-approximated by a two-level finite duration Landau-Zener problem between a midgap state and a $|\phi\rangle$ state \cite{vitanov1996landau, rubbmark1981dynamical}. In order for the transfer probability to be unity, one needs both that the sweep rate is sufficiently small, and that the frequency range of the sweep is sufficiently large. For our system, these two requirements yield
\begin{equation}
k_{\mbox{\tiny{F}}} k_0 \gtrsim 10\hspace{0.05cm} \Omega \hspace{0.05cm} |\bar{u}^{g}_k| \;\; 
\text{and} \;\; \Omega \hspace{0.05cm} |\bar{u}^{g}_k| \gtrsim \sqrt{\nu} \;\;\text{for all} \;\; k \in [-k_0,0] \hspace{0.05cm}.
\label{vacatecondition1}
\end{equation}
In Sec.~\ref{spectrum} we showed that for sufficiently strong interactions ($w \lesssim 3$, where $w \equiv (k_0/k_{\mbox{\tiny{F}}}) \exp(\pi k_{\mbox{\tiny{F}}} a_{\mbox{\tiny{1D}}} / 2)$), $|\bar{u}^{g}_k|$ is well-approximated by \smash{$|\bar{u}^{g}_k| \approx \sqrt{\xi} \hspace{0.03cm}\sech(2\xi k/k_0)$}, where $\xi = \pi w/16$ measures the spatial width of a midgap state around a soliton core. Thus we can rewrite the conditions in Eq.~\eqref{vacatecondition1} in terms of $w$ as
\begin{align}
k_{\mbox{\tiny{F}}} k_0 \gtrsim 10\hspace{0.05cm} \Omega \hspace{0.03cm} \sqrt{\xi} \;\;\text{and}\;\; \Omega \hspace{0.03cm} \sqrt{\xi} \gtrsim \sqrt{\nu} \cosh(2\xi)\hspace{0.05cm},
\label{allboundstatescondition}\\
\text{where}\;\;\xi(w) \approx \pi w/16 \hspace{0.05cm}.
\label{xifunction}
\end{align}
Note that $\Omega \hspace{0.03cm} \sqrt{\xi}$ acts as the effective coupling strength. This is sensible because the coupling strength involves the inner product of a midgap state and a plane wave, which is indeed proportional to the square root of the width $\xi$ of the midgap state.

\subsubsection{\label{pauliblocking}Eliminating bulk excitations by Pauli blocking}
As previously explained, if one starts with a Fermi sea of $|\phi\rangle$-atoms, Pauli blocking prevents any excitation from the occupied bulk modes. This requires that the Fermi energy $\epsilon_{\phi}$ is sufficiently large. However, if it is too large, one may transfer atoms from the $|\phi\rangle$ states to the vacant quasiparticle modes. Here we calculate the bounds on $\epsilon_{\phi}$. We find that the lower and upper bounds are well separated for strong enough interactions.

We can estimate a lower bound on $\epsilon_{\phi}$ by calculating the effect on the bulk mode $b^*$ with the smallest detuning from resonance, which occurs at \smash{$k_{\mbox{\tiny{F}}} + k = k_{\phi} \equiv \sqrt{2 \hspace{0.03cm} \epsilon_{\phi}}$} when $\omega = \omega_{\text{max}}$ [Fig.~\ref{cffloprotocolfig}(b)]. The finite duration Landau-Zener problem gives negligible transfer probability if
\begin{equation}
\epsilon_{\mbox{\tiny{F}}} - \epsilon_{\phi} \lesssim \epsilon_+ -  k_{\mbox{\tiny{F}}} k_0/2 - 10\hspace{0.05cm}\Omega\hspace{0.05cm} \big|\bar{u}^{b^{\hspace{-0.03cm}*}}_{k_{\phi} - k_{\mbox{\tiny{F}}}}\big| \hspace{0.08cm}.
\label{nobulktophi}
\end{equation}
Similarly, the upper bound on $\epsilon_{\phi}$ is set by requiring that no particle is transferred from a filled $|\phi\rangle$ state to an empty midgap state. The smallest detuning for such a coupling occurs at $k = k_{\phi} - k_{\mbox{\tiny{F}}}$ when $\omega = \omega_{\text{min}}$ [Fig.~\ref{cffloprotocolfig}(c)]. The transfer probability approaches zero if
\begin{equation}
\epsilon_{\mbox{\tiny{F}}} - \epsilon_{\phi} \gtrsim \epsilon_- + 3 k_{\mbox{\tiny{F}}} k_0/2 + 10\hspace{0.05cm}\Omega\hspace{0.05cm} \big|\bar{u}^{g}_{k_{\phi} - k_{\mbox{\tiny{F}}}}\big| \hspace{0.08cm}.
\label{nophitomidgap}
\end{equation}
The conditions in Eqs.~\eqref{nobulktophi} and \eqref{nophitomidgap} simplify for $w \lesssim 2$, where $\epsilon_- \approx 0$, $\epsilon_+ \approx 4 k_{\mbox{\tiny{F}}} k_0/w$, \smash{$|\bar{u}^{g}_k| \approx \sqrt{\xi} \hspace{0.03cm}\sech(2\xi k/k_0)$}, and \smash{$|\bar{u}^{b^{\hspace{-0.03cm}*}}_k| \approx |k_0/\pi k|$} for $k<\hspace{-0.05cm}-k_0/2$ (details in Sec.~\ref{spectrum}). Combining these estimates with Eq.~\eqref{allboundstatescondition} and using the inequality \smash{$k_{\mbox{\tiny{F}}} - k_{\phi} \gtrsim (\epsilon_{\mbox{\tiny{F}}} - \epsilon_{\phi})/k_{\mbox{\tiny{F}}} > 0$}, we can write
\begin{align}
\hspace{0.05cm}\Omega\hspace{0.05cm} \big|\bar{u}^{g}_{k_{\phi} - k_{\mbox{\tiny{F}}}}\big| &< \hspace{0.05cm}\Omega\hspace{0.03cm}\sqrt{\xi} \lesssim 0.1\hspace{0.05cm} k_{\mbox{\tiny{F}}} k_0\hspace{0.05cm}, \;\;\text{and}\\
\Omega\hspace{0.05cm} \big|\bar{u}^{b^{\hspace{-0.03cm}*}}_{k_{\phi} - k_{\mbox{\tiny{F}}}}\big| &\lesssim \frac{1}{\pi} \frac{\Omega\hspace{0.05cm} k_{\mbox{\tiny{F}}} k_0}{\epsilon_{\mbox{\tiny{F}}} \hspace{-0.05cm}- \hspace{-0.05cm} \epsilon_{\phi}} \lesssim \frac{0.1}{\pi \sqrt{\xi}}\frac{(k_{\mbox{\tiny{F}}} k_0)^2}{\epsilon_{\mbox{\tiny{F}}}  \hspace{-0.05cm}- \hspace{-0.05cm} \epsilon_{\phi}} \approx \frac{0.4 }{\pi^{\frac{3}{2}}\sqrt{w}}\frac{(k_{\mbox{\tiny{F}}} k_0)^2}{\epsilon_{\mbox{\tiny{F}}}  \hspace{-0.05cm}- \hspace{-0.05cm} \epsilon_{\phi}}.
\label{amplitudebounds}
\end{align}
Substituting these upper bounds into Eqs.~\eqref{nobulktophi} and \eqref{nophitomidgap}, we find that the inequalities will be satisfied if
\begin{equation}
(5/2)\hspace{0.03cm} k_{\mbox{\tiny{F}}} k_0 \lesssim \epsilon_{\mbox{\tiny{F}}} - \epsilon_{\phi} \lesssim \big(4/w - 1/2 - \sqrt{w}/\pi^{3/2}\big)\hspace{0.03cm} k_{\mbox{\tiny{F}}} k_0 \hspace{0.05cm}.
\label{ephibounds}
\end{equation}
Note that Eq.~\eqref{ephibounds} gives only sufficient, not necessary, conditions on the energy threshold $\epsilon_{\phi}$. In practice, the bounds on $\epsilon_{\phi}$ would be less stringent than in Eq.~\eqref{ephibounds}.

\subsubsection{\label{nopauliblocking}Bulk excitations without Pauli blocking}
If all $|\phi\rangle$ states are initially empty, the RF coupling will excite particles from the occupied bulk modes to these empty $|\phi\rangle$ states. Here we estimate an upper bound on the probability $P_b$ of such excitations.

The coherence factors $|\bar{u}^{j}_{k}|$ in Eq.~\eqref{HRFplus} fall off as one moves away from the Fermi point. Thus $P_b$ is maximum for the bulk mode $b_+$ which is resonantly coupled to a $|\phi\rangle$ state at the smallest magnitude of $k$, which occurs for $(k_{\mbox{\tiny{F}}} + k)^2/2 \approx \epsilon_{\mbox{\tiny{F}}} - \epsilon_+$ [Figs.~\ref{cffloprotocolfig}(b) and \ref{cffloprotocolfig}(c)], or $k \approx \sqrt{2(\epsilon_{\mbox{\tiny{F}}} - \epsilon_+)} - k_{\mbox{\tiny{F}}} \lesssim k_+ \equiv -\epsilon_+/k_{\mbox{\tiny{F}}}$. For $w \lesssim 2$, $k_+\approx 4 k_0/w$ (see Sec.~\ref{spectrum}). The corresponding coherence factor is given by \smash{$|\bar{u}^{b_+}_{k_+}| \approx |k_0/\pi k_+| \approx w/(4\pi)$}, which is linear in $w$. In contrast, the coherence factors for resonantly coupled midgap modes (see Sec.~\ref{adiabaticity}) are given by $|\bar{u}^{g}_{k}| \gtrsim \smash{\sqrt{\pi w/16}}\hspace{0.02cm} \sech(\pi w/8) \sim \mathcal{O}(\sqrt{w})$ for small $w$. Hence the coherence factors for bulk excitations fall off much faster with stronger interactions (smaller $w$), which means one can tune the coupling strength $\Omega$ so that the RF sweep is adiabatic for midgap modes, but diabatic for bulk modes.

In particular, at the lower bound of the coupling strength for adiabaticity in Eq.~\eqref{allboundstatescondition}, a Landau-Zener analysis for the bulk mode $b_+$ gives
\begin{equation}
P_b \lesssim 1 - e^{-2 \pi \big|\bar{u}^{b_+}_{k_+}\big|^2 \Omega^2 / \nu} \approx 1 - e^{-(2 w / \pi^2)\hspace{0.01cm} \cosh^2 \hspace{-0.01cm}(\pi w /8)} \hspace{0.05cm},
\label{Pb}
\end{equation}
which falls toward zero as interactions are increased. For $w = 2$, $P_b \lesssim 0.5$, and for $w=1$, $P_b \lesssim 0.2$. Thus even without Pauli blocking, one excites a small fraction of the bulk modes at strong enough interactions.

\subsubsection{\label{instability}Condition for dynamical stability}
In \cite{[{}] [{ ({\it accepted in Phys. Rev. Lett.}). }]dutta2016collective} we showed that a balanced soliton train has dynamical instabilities toward a uniform superfluid phase. The instability consists of neighboring solitons approaching one another and annihilating after a characteristic lifetime set by the maximum instability rate $\eta_{\text{max}}$. For our protocol to work properly, the sweep duration $\tau_{\text{sw}}$ must be short compared to this lifetime, i.e., $\tau_{\text{sw}} \ll \eta_{\text{max}}^{-1}$, as otherwise the soliton train would decay before the sweep is completed. In \cite{[{}] [{ ({\it accepted in Phys. Rev. Lett.}). }]dutta2016collective} we found an upper bound on $\eta_{\text{max}}$ in the full BdG dynamics, $\eta_{\text{max}} \lesssim 2 \sqrt{\epsilon_+ \epsilon_-}$. For $w \lesssim 2$, this upper bound can be expressed as (see Sec.~\ref{spectrum})
\begin{align}
\eta_{\text{max}} \lesssim 2 \sqrt{\epsilon_+ \epsilon_-} \approx k_{\mbox{\tiny{F}}} k_0 \hspace{0.03cm} f(w) \hspace{0.05cm},
\label{etamax}\\
\text{where}\;\; f(w) \approx 16 \hspace{0.05cm} w^{-1} e^{-2\pi/w}\hspace{0.05cm}.
\label{ffunction}
\end{align}
Note that $\eta_{\text{max}}$ decreases sharply with $w$, as stronger interactions stabilize the soliton train. The sweep duration is given by $\tau_{\text{sw}} = 2 k_{\mbox{\tiny{F}}} k_0 / \nu$. Hence, the condition $\tau_{\text{sw}} \ll \eta_{\text{max}}^{-1}$ will be satisfied if
\begin{equation}
1/f(w) \gg 2\hspace{0.03cm} (k_{\mbox{\tiny{F}}} k_0)^2 / \nu \hspace{0.05cm}.
\label{instabilitycondition}
\end{equation}
Note that Eq.~\eqref{instabilitycondition} is again a sufficient condition, not a necessary one.

\subsubsection{\label{stronginteractions}Implication for interaction strength}
Combining the adiabaticity requirements in Eq.~\eqref{allboundstatescondition} and the stability condition in Eq.~\eqref{instabilitycondition}, we obtain
\begin{equation}\frac{1}{f(w)} \gg\frac{2(k_{\mbox{\tiny{F}}} k_0)^2}{\nu} \gtrsim 200\hspace{0.05cm}\frac{\Omega^2 \hspace{0.02cm}\xi(w)}{\nu} \gtrsim 200 \cosh^2(2\xi(w))
\label{overallinequality}
\end{equation}
[Recall, $w$ parametrizes the interaction strength, $f(w)$ is given by Eq.~\eqref{ffunction}, and $\xi(w)$ is given by Eq.~\eqref{xifunction}]. To satisfy this inequality, one must have $1/f(w) \gg 200 \cosh^2(2\xi(w))$, which occurs for $w < 3/4$, i.e., in the strongly interacting regime. Quantitative calculations in this regime may require going beyond the Andreev approximation. Nevertheless, our estimates should be robust. Firstly, the procedure itself rests on very generic features which do not depend on the specifics of the model, such as (i) the principle of Rapid Adiabatic Passage to transfer particles between two states \cite{vitanov2001laser, moller2008quantum, rangelov2005stark, malinovsky2001general, vitanov1996landau, rubbmark1981dynamical}, (ii) a separation of energy scales between the localized and bulk quasiparticle modes, and (iii) symmetry properties of a soliton train. Therefore Eqs.~\eqref{allboundstatescondition}, \eqref{nobulktophi}, \eqref{nophitomidgap}, and \eqref{instabilitycondition} remain valid in the full model. We have only invoked the Andreev approximation in writing down expressions for $\xi(w)$ and $f(w)$ in Eqs.~\eqref{xifunction} and \eqref{ffunction}, and in estimating the bounds in Eqs.~\eqref{ephibounds} and \eqref{Pb}. By numerically solving the full BdG equations, we find good agreement with these estimates at strong interactions. Further, as we discussed earlier, stronger interactions yield a large bulk gap, which increases the critical temperature of the superfluid \cite{mertsching1981incommensurate, machida1984superconductivity, [{}] [{ [\href{http://www.jetp.ac.ru/cgi-bin/e/index/e/66/2/p422?a=list}{Sov. Phys. JETP {\bf 66}, 422 (1987)}]. }]buzdin1987nonuniform, [{}] [{ [\href{http://www.jetp.ac.ru/cgi-bin/e/index/e/58/2/p428?a=list}{Sov. Phys. JETP {\bf 58}, 428 (1983)}]; }]buzdin1983phase}, thus reducing the effect of thermal fluctuations which we have ignored. Hence, our protocol will have a high fidelity in the strongly interacting regime. Note that experimentalists routinely tune the atomic interactions from very small to very large values using a Feshbach resonance \cite{chin2010feshbach}.

\subsubsection{\label{typicalparameters}Typical experimental parameters}
As a specific example, suppose we would like to create a C-FFLO state where adjacent domain walls are separated by $\pi/k_0 \sim 10\;\mu$m. This lengthscale is compatible with phase imprinting, where achievable lengthscales are ultimately limited by diffraction. We consider the parameters in \cite{liao2010spin} where $^{6}$Li atoms were trapped in quasi-1D tubes with $\epsilon_{\mbox{\tiny{F}}} = 1.2\;\mu$K, or $k_{\mbox{\tiny{F}}} = 5.4 \times 10^6 \;\mbox{m}^{-1}$. Then a $10\;\mu$m soliton spacing corresponds to $k_0/k_{\mbox{\tiny{F}}} \approx 0.05$. To ensure the soliton train is stable thoughout the sweep, we require $\eta_{\text{max}} \tau_{\text{sw}} \approx 1/20$. Then Eq.~\eqref{overallinequality} gives
\begin{equation}
\frac{1}{f(w)} \gg\frac{40(k_{\mbox{\tiny{F}}} k_0)^2}{\nu} \gtrsim 4000\hspace{0.05cm}\frac{\Omega^2 \hspace{0.02cm}\xi(w)}{\nu} \gtrsim 4000 \cosh^2(2\xi(w))\hspace{0.03cm}.
\label{example}
\end{equation}
Comparing the first and last terms, we get $w \lesssim 0.54$ or $k_{\mbox{\tiny{F}}} a_{\mbox{\tiny{1D}}}\lesssim 1.5$, which could be set by tuning a magnetic field around a Feshbach resonance \cite{chin2010feshbach}. For comparison, in \cite{liao2010spin} $k_{\mbox{\tiny{F}}} a_{\mbox{\tiny{1D}}} \approx 0.6$. For $k_{\mbox{\tiny{F}}} a_{\mbox{\tiny{1D}}} = 1.5$, the instability rate is $\eta_{\text{max}} \lesssim  k_{\mbox{\tiny{F}}} k_0 f(w) \approx 3.8 \;\mbox{s}^{-1}$, which gives a sweep duration $\tau_{\text{sw}} \sim 1/(20\hspace{0.03cm} \eta_{\text{max}}) \approx 13\;$ms. During this interval the frequency is to be varied over a range $\Delta\omega = 2 k_{\mbox{\tiny{F}}} k_0 \approx 5$ kHz, at a rate $\nu \approx 3.8 \times 10^5$ Hz/s. Equating the middle terms in Eq.~\eqref{example} yields a Rabi frequency \smash{$\Omega \sim k_{\mbox{\tiny{F}}} k_0 / (10 \sqrt{\xi}\hspace{0.04cm}) \approx 0.77$} kHz. To suppress unwanted quasiparticle excitations, one can fill up all $|\phi\rangle$ states with energy below $\epsilon_{\phi}$ where, from Eq.~\eqref{ephibounds}, 0.4~$\mu$K $\lesssim \epsilon_{\phi} \lesssim$ \smash{0.9 $\mu$K}. These numbers are well within reach of present-day experiments. Even if one starts with no $|\phi\rangle$-atoms, from Eq.~\eqref{Pb} we find that the sweep will only excite less than 11\% of the bulk modes.

\section{\label{discussion}Summary and Outlook}
We have described a simple experimental protocol to engineer long-lived FFLO states in a two-component gas of cold fermionic atoms loaded in a quasi-1D trap. The protocol consists of first preparing a train of domain walls in a balanced superfluid by phase imprinting, then using a radio-frequency sweep to selectively transfer the spin-$\uparrow$ atoms near the domain walls to a third noninteracting spin state $|\phi\rangle$, leaving behind an FFLO state with exactly one unpaired fermion per domain wall. Prior work has shown that this engineered configuration is stable \cite{[{}] [{ ({\it accepted in Phys. Rev. Lett.}). }]dutta2016collective}. By analyzing the different limiting factors, we have shown that the protocol can be implemented with high fidelity for sufficiently strong interactions which are readily attainable in current experimental set-ups. It provides a route to directly produce FFLO states in experiments in a controlled manner and study their properties. Such a direct approach complements the thermodynamic search of the exotic state and contributes to the larger goal of engineering many-body quantum states.

In describing the protocol, we have analyzed the case where the frequency is swept over an interval $\Delta\omega = 2 k_{\mbox{\tiny{F}}} k_0$ [Figs.~\ref{cffloprotocolfig}(b) and \ref{cffloprotocolfig}(c)], as this is the shortest sweep which is expected to transfer all of the particles from the localized modes. One can also sweep over larger frequency intervals, but the analysis would have to be repeated to ensure the broader sweep did not excite bulk modes.

Our procedure yields an FFLO state in the presence of a gas of $|\phi\rangle$ atoms. Since the $|\phi\rangle$-atoms are very weakly interacting, they should not affect the dynamics of the soliton train. Alternatively, one could remove all $|\phi\rangle$-atoms after the sweep by a resonant optical pulse \cite{du2008observation}.

The generated FFLO state can be probed using a variety of techniques that have been proposed in the literature \cite{lutchyn2011spectroscopy, edge2009signature, edge2010collective, lu2012expansion, hu2011josephson, swanson2012proposal, bakhtiari2008spectral, mizushima2005direct, korolyuk2010probing, kajala2011expansion, gritsev2008interferometric, roscilde2009quantum, luscher2008fulde}. For example, one can excite collective modes by ramping to a different interaction strength. If the ramp is fast compared to the bulk gap $\epsilon_+ \approx 4 k_{\mbox{\tiny{F}}} k_0/w$, the domain walls will not have time to adjust their shape, which will excite a novel collective mode where the width of each domain wall oscillates in time \cite{[{}] [{ ({\it accepted in Phys. Rev. Lett.}). }]dutta2016collective}. The collective modes could be detected using spectroscopic or imaging techniques \cite{ku2014motion, ku2016cascade, lutchyn2011spectroscopy, edge2009signature, edge2010collective, torma2016physics, endres2012higgs}.

Our protocol could be generalized to create incommensurate FFLO states which have less than one excess fermion per soliton, for example, by sweeping over smaller frequency intervals such that only a fraction of the midgap modes are resonantly driven during a sweep. However, since the midgap modes are contiguous in energy, it would be more challenging to control the number of unpaired fermions per soliton.

Finally, a recent study has shown that domain walls are also stabilized in 3D when filled with unpaired fermions \cite{reichl2017core}, which could offer ways of extending our protocol to higher dimensions.

\section{\label{acknowledgments}Acknowledgments}
This material is based upon work supported by the National Science Foundation Grant PHY-1508300.


%

\end{document}